\documentclass[journal]{IEEEtran}

\usepackage{amsmath,amssymb,amsfonts}
\usepackage{graphicx}
\usepackage{cite}
\usepackage{xcolor}
\usepackage{booktabs}
\usepackage{multirow}
\usepackage{siunitx}       % for \SI{}{} units
\usepackage{physics}       % for \qty, bra-ket notation
\usepackage{hyperref}
\usepackage{cleveref}
\usepackage{subcaption}
\usepackage{algorithm}
\usepackage{algorithmic}
\usepackage{orcidlink}

\newcommand{\Tone}{T_1}

\newcommand{\Tsys}{T_\mathrm{sys}}
\newcommand{\Tn}{T_N}
\newcommand{\chidisp}{\chi}
\newcommand{\sigzero}{\sigma_0}
\newcommand{\sigone}{\sigma_1}
\newcommand{\fid}{\mathcal{F}}
\newcommand{\snr}{\mathrm{SNR}}

\begin{document}

\title{Qubit-Based Benchmarking of InP HEMT LNAs:
       Readout Fidelity Versus Power Consumption}

% Author block — adjust as needed
\author{
  Junjie~Li\orcidlink{0000-0002-0081-3365}
  \thanks{Manuscript received \today.}
  \thanks{J. Li is with the Chalmers Next Labs AB, WACQT Quantum Technology Testbed, SE-41296 Gothenburg, Sweden (e-mail: junjie.li@chalmersnextlabs.se).}
}

\markboth{Journal,~Vol.~XX,~No.~X,~2026}%
{Author \MakeLowercase{\textit{et al.}}: HEMT Benchmarking via Qubit Readout Fidelity}

\maketitle

% ---------------------------------------------------------------
\begin{abstract}
High-fidelity single-shot readout of superconducting qubits is
essential for fault-tolerant quantum computation.
Without a quantum-limited amplifier such as a Josephson parametric
amplifier, the cryogenic high-electron-mobility-transistor (HEMT)
low-noise amplifier (LNA) at the 4 K stage is the dominant
noise source in the readout chain.
HEMT LNAs are conventionally characterized by Y-factor measurements
of the noise temperature $T_N$, independently of the quantum
measurement they ultimately serve.
Here we present a qubit-in-the-loop benchmarking method.
The method uses the two-dimensional IQ histogram of single-shot
readout as a direct figure of merit.
We apply it to three state-of-the-art cryogenic InP HEMT LNAs with 53\%, 60\%,
and 70\% channel indium content, together with a commercial
reference amplifier.
The 60\% and 70\% devices have similar $T_N$ in Y-factor measurements, both lower than the 53\%
device. The 70\% device has the highest gain, but qubit readout SNR and assignment fidelity $\mathcal{F}_a$ instead peak at 60\%.
Using this method, we map $\mathcal{F}_a$ against HEMT LNA dc power
consumption for each device.
$\mathcal{F}_a$ does not simply saturate with increasing power.
For several devices it declines above a device-dependent power well
below typical operating points.
For the 60\% device, which has the highest overall $\mathcal{F}_a$, about
1~mW is needed to maintain $\mathcal{F}_a > 85\%$.
Across all four devices, $\mathcal{F}_a > 80\%$ is reached with as little as
about 0.3~mW, roughly an order of magnitude below typical
operating points.
These results establish a qubit-referenced benchmarking methodology
and provide practical guidance for amplifier selection and bias in
power-constrained multi-qubit readout systems.
\end{abstract}


% ===============================================================
\section{Introduction}
\label{sec:intro}
% ===============================================================

Superconducting qubits processed via circuit quantum electrodynamics
(cQED) are among the leading platforms for fault-tolerant quantum
computing~\cite{blais2021cqed, krantz2019, devoret2013outlook}.
A critical operation in any quantum processor is the high-fidelity
readout of individual qubit states, which is required for quantum
error correction protocols~\cite{fowler2012surface} and mid-circuit
measurements~\cite{walter2017rapid}.

In the dispersive readout scheme, the qubit state is encoded in the
phase of a microwave probe tone reflected from a coupled resonator.
The resulting signal is extremely weak, typically on the order of
\SI{-120}{dBm} at the output of the readout resonator \cite{krantz2019}. The signal must be
amplified before detection.
The amplification chain typically consists of a Josephson parametric
amplifier (JPA) or a Josephson traveling-wave parametric amplifier
(TWPA)~\cite{macklin2015twpa} at millikelvin temperatures followed by a cryogenic
 high-electron-mobility-transistor (HEMT)
low-noise amplifier (LNA) at the \SI{4}{K} stage, and room-temperature amplifiers
thereafter~\cite{Arute2019}.
When a quantum-limited JPA/TWPA is absent, the HEMT LNA becomes the dominant
noise source, adding on the order of 10 noise quanta at typical
readout frequencies~\cite{hover2014slug}, which severely limits
single-shot readout fidelity.

Even in systems equipped with a JPA/TWPA, the HEMT LNA plays a critical
secondary role: its noise contribution enters the Friis chain
divided by the JPA/TWPA gain, and its dc power dissipation directly
loads the \SI{4}{K} cooling stage, which is a scarce resource
in large-scale processors~\cite{krinner2019cryogenic}.
As quantum processors scale toward hundreds and thousands of qubits, the power budget of the
\SI{4}{K} stage becomes a fundamental engineering constraint.

Existing HEMT LNA characterization methods, including Y-factor noise
temperature measurements~\cite{pozar2011microwave}, evaluate
amplifier performance using microwave test equipment in isolation
from the quantum measurement chain.
While these measurements are essential for device selection and
qualification~\cite{cha2023subMW}, they do not directly capture
how amplifier noise translates into qubit state discrimination
errors in a realistic readout scenario.
In particular, Y-factor measurements cannot account for effects
such as the amplifier's impact on qubit coherence through thermal
photon emission, or the interaction between amplifier noise and the
nonlinear dynamics of the readout resonator.

In this work, we bridge this gap by demonstrating a
\emph{qubit-in-the-loop} benchmarking methodology in which the
two-dimensional IQ histogram of single-shot readout measurements
serves as the primary figure of merit for cryogenic HEMT evaluation.
The three InP HEMT LNAs with channel indium content of 53\%, 60\% and 70\% were previously characterized in~\cite{ouryFactor} via Y-factor methods; here we
extend it to the quantum measurement domain and establish a direct
link between amplifier noise temperature, IQ cloud statistics, and
assignment fidelity.

% ===============================================================
\section{Background}
\label{sec:background}
% ===============================================================

%\subsection{Dispersive Readout and Amplifier Noise}
%\label{sec:dispersive}

In cQED, a transmon qubit with transition frequency $\omega_q$ is
coupled with strength $g$ to a readout resonator at frequency
$\omega_r$~\cite{blais2021cqed, koch2007transmon}.
In the dispersive limit ($|\omega_q - \omega_r| \gg g$), the
qubit-state-dependent resonator frequency shift is
\begin{equation}
  \omega_r^{(0,1)} = \omega_r \pm \chidisp,
  \quad
  \chidisp \approx \frac{g^2}{\omega_q - \omega_r},
\label{eq:dispersive_shift}
\end{equation}
where $\chidisp$ is the dispersive shift.
Probing the resonator with a coherent tone and performing homodyne
detection yields a complex IQ voltage
\begin{equation}
  V_{IQ} = I + iQ = \frac{A}{2} e^{i\phi_{0,1}},
\end{equation}
where the phase $\phi_{0,1}$ depends on the qubit state. A single-shot readout measurement yields one point $V_{IQ}$ in the two-dimensional IQ plane. Repeating this measurement many times for
a fixed qubit state does not return the same point every time:
noise added throughout the amplification chain scatters the
individual shots into a roughly Gaussian-distributed cloud, or
\emph{blob}. Preparing the qubit in $\ket{0}$, $\ket{1}$ and repeating single-shot measurements yields two blobs in the IQ plane, centered at $\mu_0$ and $\mu_1$ with widths $\sigzero$ and $\sigone$, respectively. The centroid position encodes the qubit state, while the width reflects the noise added by the measurement chain.

The signal-to-noise
ratio of the measurement can be expressed by~\cite{walter2017rapid}
\begin{equation}
  \snr = \frac{d}{\bar{\sigma}}
       = \frac{|\mu_1 - \mu_0|}{(\sigzero + \sigone)/2}.
\label{eq:snr}
\end{equation}

In the ideal case where $\sigzero=\sigone=\sigma$, the assignment
fidelity is the standard analytic form~\cite{mallet2009singleshot}:
\begin{equation}
   \fid_\mathrm{id} = \frac{1}{2}\left[1+\mathrm{erf}\!\left(\sqrt{\frac{\snr^2}{8}}\right)\right].
\label{eq:fidelity_ideal}
\end{equation}

In practice, however, $\sigone>\sigzero$. $\sigzero$ is set
essentially by the amplifier noise referred to the input \cite{walter2017rapid,pozar2011microwave}:

\begin{equation}
  \sigma_\mathrm{0}^2 \propto G k_B \Tsys \cdot \Delta f,
\label{eq:sigma_amp}
\end{equation}

\noindent where $G$ is the chain's total voltage gain, $k_B$ is the Boltzmann constant, $\Delta f$ is the measurement bandwidth and
$\Tsys$ is the system noise temperature referred to the amplifier input. By the Friis formula~\cite{pozar2011microwave}:
\begin{equation}
  \Tsys = T_1 + \frac{T_2}{G_1} + \frac{T_3}{G_1 G_2} + \cdots,
\label{eq:friis}
\end{equation}
\noindent with $T_i$ and $G_i$ being the noise temperature and gain of each
stage. Since the HEMT LNA is the first gain stage in the absence of a
JPA/TWPA, its noise temperature dominates $\Tsys$.

$\sigone$ contains this same amplifier-noise contribution, plus
additional broadening from qubit decay affected by the relaxation time $\Tone$ during the readout window ~\cite{gambetta2007quantum}. And thermal population with prepare $\ket{0}$ shot that starts thermally in $\ket{1}$ also affects $\sigone$. Thus the actual assignment fidelity $\fid_a$ is computed empirically from the
classifier's outcomes without assuming Gaussian statistics or equal
widths \cite{chen2023readout}:
\begin{equation}
  \fid_a = 1-\frac{P(1|0)+P(0|1)}{2},
\label{eq:fidelity_actual}
\end{equation}
where $P(i|j)$ is the probability that a shot prepared in $\ket{j}$
is assigned to $\ket{i}$ by the classifier. $\fid_a$ is the quantity reported throughout this work.

%\subsection{IQ Histogram and Assignment Fidelity}
%\label{sec:fidelity_theory}

%\subsection{HEMT Bias and Power Consumption}
%\label{sec:hemt_bias}

% ===============================================================
\section{Experimental Setup}
\label{sec:setup}
% ===============================================================

\begin{figure}[h]
  \centering
  \includegraphics[width=0.7\columnwidth]{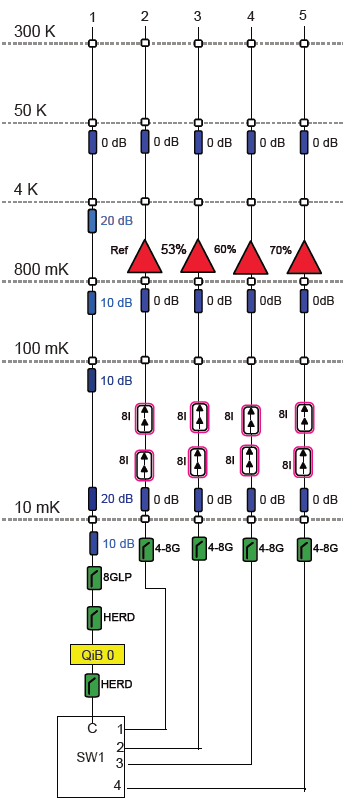}
  \caption{Fridge wiring diagram. Column 1 is the input line;
           columns 2--5 are the four HEMT LNA output lines, selected
           by switch SW1 at the mixing-chamber stage.
           Total attenuation (blue rectangles) on the
           input line from 300\,K to the sample
           is \SI{70}{dB}.
           The HEMT LNAs (Red triangles) are mounted at the
           \SI{4}{K} stage.
           4-8 GHz isolators (8I) and filters (green boxes, a 4--8\,GHz bandpass filter on
           each output line; a 8 Ghz low-pass filter (8GLP) and two
           IR-blocking filter (HERD) on the input line).
           The yellow box (QiB\,0) is the qubit chip.
          }
  \label{fig:setup}
\end{figure}

The experimental setup is shown in Fig.~\ref{fig:setup}.
A transmon qubit~\cite{koch2007transmon} is coupled to a coplanar
waveguide readout resonator and housed at the \SI{10}{mK} stage
of a dilution refrigerator.
Table~\ref{tab:qubit_params} summarizes the qubit and resonator
parameters used throughout this study. To compare the four HEMT LNAs without thermal cycling the
cryostat, each device is wired into its own output
chain (columns 2--5 in Fig.~\ref{fig:setup}), and a cryogenic
Radiall switch (SW1) mounted at the mixing-chamber stage. It can select which output chain is connected to the qubit's readout line for a given measurement. The qubit, the input line and
the room-temperature electronics are identical to all four output lines. Thus only the HEMT LNA under test differs between measurements.
All HEMT LNA comparison experiments are performed at a fixed readout
tone power at the device input and an integration
time of $T_\mathrm{int} = 3~\mu$s.

\begin{table}[h]
  \caption{Qubit and Resonator Parameters}
  \label{tab:qubit_params}
  \centering
  \begin{tabular}{lcc}
    \toprule
    Parameter & Symbol & Value \\
    \midrule
    Qubit frequency        & $\omega_q / 2\pi$         & 3.782~GHz \\
    Resonator frequency    & $\omega_r / 2\pi$         & 5.938~GHz \\
    Dispersive shift       & $\chidisp / 2\pi$         & 0.043~MHz \\
    Resonator linewidth    & $\kappa / 2\pi$           & 0.15~MHz \\
    Energy relaxation time & $\Tone$                   & 120 ~$\mu$s \\
    Ramsey dephasing time  & $T_2^*$                   & 4.5~$\mu$s \\
    Echo dephasing time    & $T_2^\mathrm{echo}$       & 188~$\mu$s \\
    \bottomrule
  \end{tabular}
\end{table}

The three state-of-the-art InP HEMT LNAs (53\%, 60\%, and 70\% channel
indium content, respectively) listed in Table~\ref{tab:hemt_params} were characterized via Y-factor
measurements in~\cite{ouryFactor}; the reference commercial device is a Low
Noise Factory LNF-LNC4\_8G, whose noise temperature is the
manufacturer-specified typical value~\cite{lnf_datasheet}.
All four devices are compared at the readout frequency of 5.9 GHz and a common dc power consumption $P_\mathrm{dc}$ of 7.8 mW:

\begin{equation}
  P_\mathrm{dc} = V_{DS} \cdot I_D
\label{eq:pdc}
\end{equation}

\noindent with bias of $V_{DS}=\SI{0.71}{V}$, $I_D=\SI{11}{mA}$ for 53\%, 60\%, and 70\% InP HEMT LNAs and $V_{DS}=\SI{0.6}{V}$, $I_D=\SI{13}{mA}$ for the reference commercial device.

InP HEMT LNAs are typically operated at a drain current $I_D$ set
to minimize noise temperature $\Tn$~\cite{cha2023subMW}.
However, $\Tn(I_D)$ is a shallow function near its optimum, while
$P_\mathrm{dc}$ varies approximately linearly with $I_D$ \cite{ouryFactor}.
This creates an optimization opportunity to reduce $P_\mathrm{dc}$ with only a modest $\Tn$ penalty ~\cite{cha2023subMW}.

\begin{table}[!t]
  \caption{HEMT Devices Under Test at 5.9 GHz}
  \label{tab:hemt_params}
  \centering
  \resizebox{\columnwidth}{!}{%
  \begin{tabular}{lcccc}
    \toprule
    Device & $\Tn$ (K) & 
    $Gain $ (dB) & $I_D$ (mA)
           & $P_\mathrm{dc}$ (mW) \\
    \midrule
    53\%             & 1.4   & 41           & 11 & 7.8 \\
    60\%             & 1.1     & 41         & 11 & 7.8 \\
    70\%             & 1.1    &  45          & 11 & 7.8 \\
    Ref (LNF) & 1.5\textsuperscript{a} & 39  & 13 & 7.8 \\
    \bottomrule
  \end{tabular}}
  \vspace{2pt}
  {\footnotesize\textsuperscript{a}~Manufacturer typical average value in 4-8 GHz~\cite{lnf_datasheet}, not measured by the author.}
\end{table}

% ===============================================================
\section{Qubit IQ Histogram Characterization}
\label{sec:baseline}
% ===============================================================

Before each HEMT LNA swap or bias change, we perform a complete qubit
characterization sequence to verify that the qubit parameters
in Table~\ref{tab:qubit_params} remain stable. $\Tone$ was found to be stable to within 5\%
across all HEMT LNA configurations, confirming that the qubit is not
significantly perturbed by HEMT LNA swaps or moderate bias changes.
As a conservative criterion, any measurement session in which a parameter drifted by more than
10\% was excluded from the analysis.

%\subsection{Thermal Population Measurement}
%\label{sec:thermal}

% ===============================================================
%\section{IQ Histogram Analysis Method}
%\label{sec:iq_method}
% ===============================================================

%\subsection{Data Acquisition}

% ===============================================================
\subsection{HEMT LNA Comparison at Nominal Bias}
\label{sec:comparison}
% ===============================================================

\begin{figure}[h]
  \centering
  \includegraphics[width=\columnwidth]{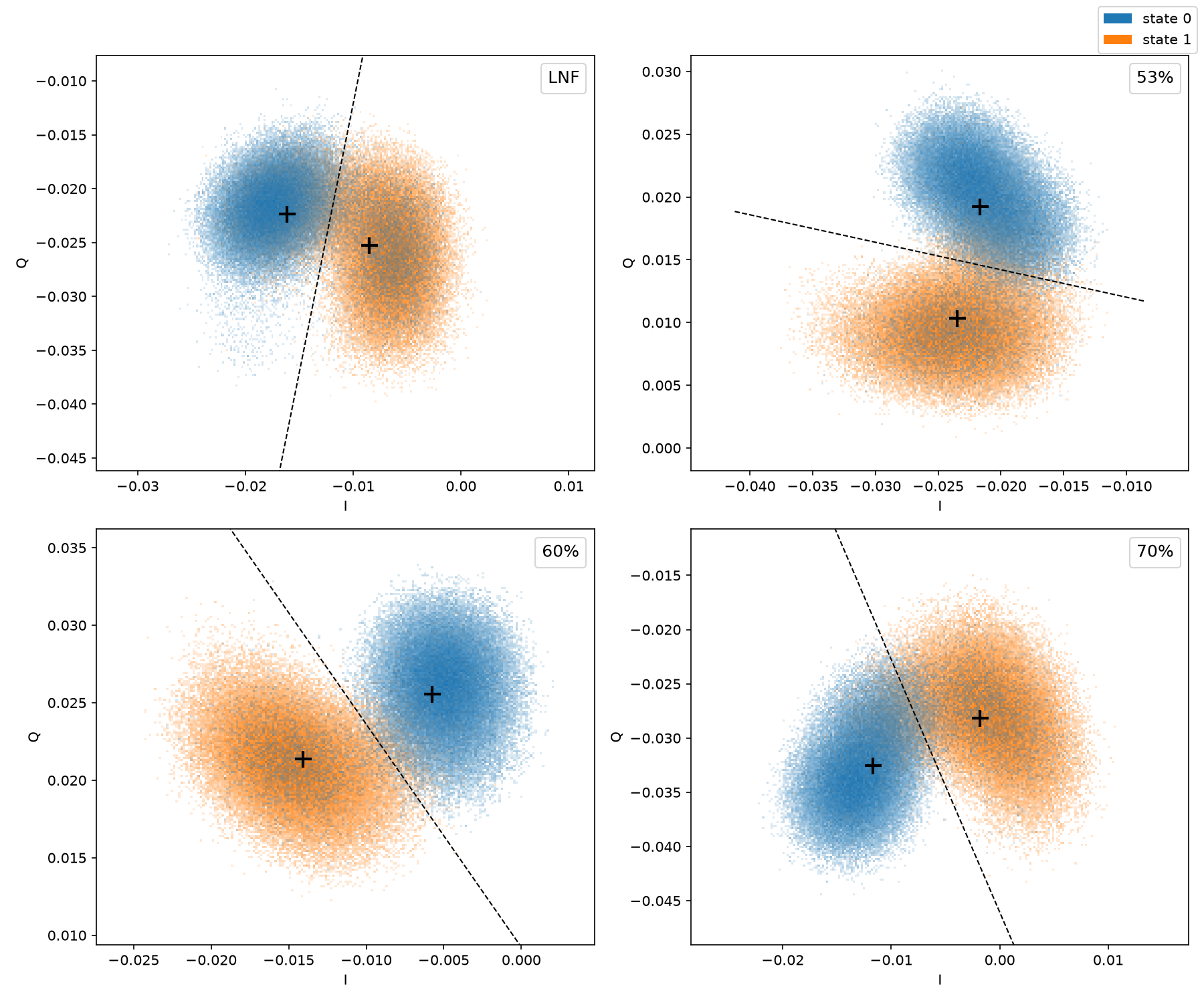}
  \caption{Two-dimensional IQ histograms for each HEMT LNA at the nominal bias with $\sim\!200{,}000$ single shots. The two clouds correspond to the qubit prepared in $\ket{0}$ (blue) and $\ket{1}$ (yellow); centroids are marked with a cross and the fitted discrimination boundary with a dashed line.
The discrimination-axis orientation differs between devices because it is set independently by each device's own IQ calibration.}
  \label{fig:iq_histograms}
\end{figure}

Fig.~\ref{fig:iq_histograms} shows the IQ histograms for all
HEMT LNAs at the nominal bias ($V_{DS}=\SI{0.6}{V}$, $I_D=\SI{13}{mA}$). For each HEMT LNA, we acquire $N = 200{,}000$ single-shot IQ measurements with the qubit prepared in $\ket{0}$, and an equal number with the qubit prepared in $\ket{1}$ (via a calibrated $\pi$ pulse). All shots are acquired with a fixed readout power and integration time to isolate the effect of the HEMT LNAs.The cloud separation $d$ is comparable across devices, confirming
that the signal (qubit + resonator) is essentially unchanged by
which HEMT LNA is selected; the blob widths, and hence the
SNR and assignment fidelity, are what differ between devices.

\begin{figure}[h]
  \centering
  \includegraphics[width=\columnwidth]{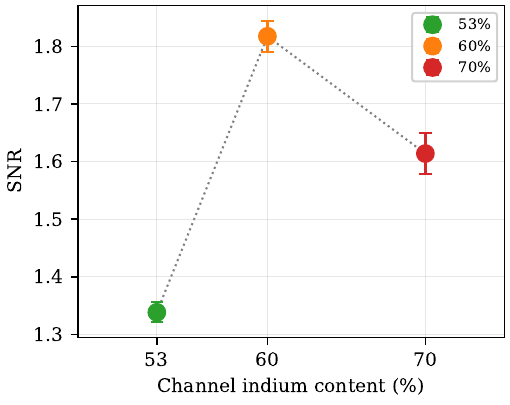}
  \caption{Readout SNR versus HEMT channel indium
           content at nominal bias. Error bars are the standard error of the mean
           over acquisition data.}
  \label{fig:snr_indium}
\end{figure}

Table~\ref{tab:comparison} summarizes the measured IQ statistics
and $\fid_a$ at the nominal bias for all devices. Fig.~\ref{fig:snr_indium} plots the same SNR values directly
against channel indium content.
The relationship is non-monotonic: the 60\% device gives the
highest SNR, while both the 53\% and 70\% devices underperform relative to it.
This optimum channel indium content of 60\% among the compositions tested here is consistent with the drain noise temperature $T_d$ reported in~\cite{ouryFactor}, though not with the overall $\Tn$ ranking.

\begin{table}[h]
  \caption{SNR and $\fid_a$ at Nominal Bias}
  \label{tab:comparison}
  \centering
  \begin{tabular}{lccc}
    \toprule
    Device & SNR & $\fid_a$ (\%)  \\
    \midrule
    53\%                   & 1.34 & 82.1  \\
    60\%                   & 1.82 & 87.2 \\
    70\%                   & 1.61 & 84.9  \\
    Commercial (LNF)       & 1.57 & 83.8  \\
    \bottomrule
  \end{tabular}
\end{table}

% ===============================================================
\subsection{Bias Optimization: Fidelity Versus Power Consumption}
\label{sec:bias_sweep}
% ===============================================================

%\subsection{Measurement Protocol}

For each HEMT LNA, we sweep the drain current $I_D$ from
\SI{1.5}{mA} to \SI{15}{mA}, with the corresponding dc power consumption $P_\mathrm{dc}$ ranging from \SI{75}{\micro\watt} to \SI{10.5}{mW}. At each bias point, we acquire $N = 10{,}000$ IQ shots each for preparation $\ket{0}$ and $\ket{1}$.

%\subsection{Fidelity--Power Trade-off}

\begin{figure}[h]
  \centering
  \includegraphics[width=\columnwidth]{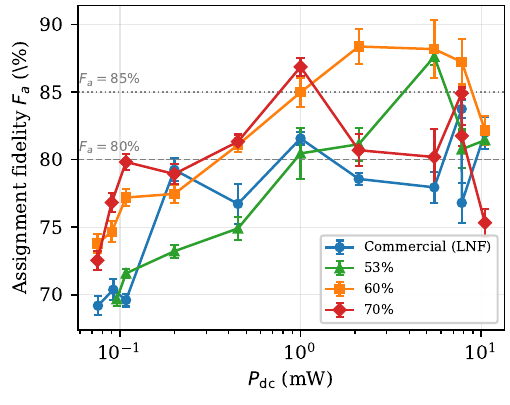}
  \caption{$\fid_a$ versus HEMT LNA dc power consumption
           $P_\mathrm{dc}$ for each device.
           Horizontal dashed lines indicate fidelity thresholds of
           80\% and 85\%.}
  \label{fig:fidelity_power}
\end{figure}

\begin{figure}[!t]
  \centering
  \includegraphics[width=\columnwidth]{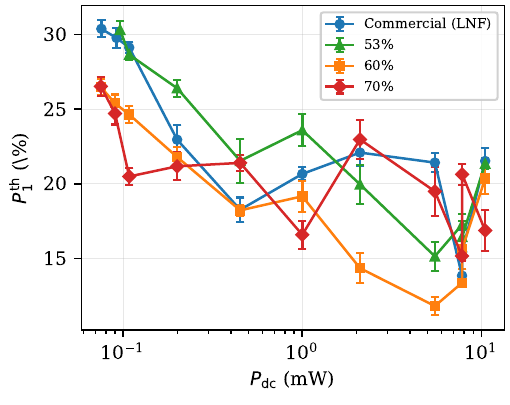}
  \caption{Qubit thermal population $P_1^\mathrm{th}$ as a function
           of $P_\mathrm{dc}$ for each
           device. Error bars are the standard error of the mean.}
  \label{fig:thermal}
\end{figure}

Fig.~\ref{fig:fidelity_power} shows the $\fid_a$
as a function of $P_\mathrm{dc}$ for each HEMT. $\fid_a$ rises quickly with bias below $\sim\SI{1}{mW}$. It fluctuates around the saturation value in the low-mW range, and for several devices decreases again above $\sim\SI{5.5}{mW}$, tracking the thermal population $P_1^\mathrm{th}$ upturn shown in Fig.~\ref{fig:thermal}. The 60\% device exceeds the commercial reference by an average of $\sim\!5$ \% across the sweep, with the largest, statistically significant gaps ($\gtrsim\!4$--10 \%) in the $\SI{0.45}{}$--$\SI{7.8}{mW}$ range.

%\subsection{Minimum Power at Target Fidelity}

Table~\ref{tab:pmin} summarizes the minimum dc power $P_\mathrm{dc}^\mathrm{min}$
required to achieve fidelity thresholds of 80\% and 85\% for each
device, obtained by interpolating the curves in
Fig.~\ref{fig:fidelity_power}.
The 70\% and 60\% devices reach $\fid>80\%$ at the lowest power
($0.29$ and $0.35$~mW respectively), a factor of roughly $22$--$27$ below
the $\SI{7.8}{mW}$ nominal bias point; the commercial device does
not reach $\fid>85\%$ anywhere in the tested range.

\begin{table}[h]
  \caption{Minimum Power for Target Fidelity}
  \label{tab:pmin}
  \centering
  \resizebox{\columnwidth}{!}{%
  \begin{tabular}{lccc}
    \toprule
    Device & $P_\mathrm{dc}^\mathrm{min}(\fid{>}80\%)$
           & $P_\mathrm{dc}^\mathrm{min}(\fid{>}85\%)$
           & $P_\mathrm{dc}^\mathrm{nom}$ \\
    \midrule
    53\%             & 0.94~mW & 3.73~mW     & 7.8~mW \\
    60\%             & 0.35~mW & 1.01~mW     & 7.8~mW \\
    70\%             & 0.29~mW & 0.77~mW     & 7.8~mW \\
    Commercial (LNF) & 0.77~mW & not reached & 7.8~mW \\
    \bottomrule
  \end{tabular}}
\end{table}

\begin{figure}[!t]
  \centering
  \includegraphics[width=\columnwidth]{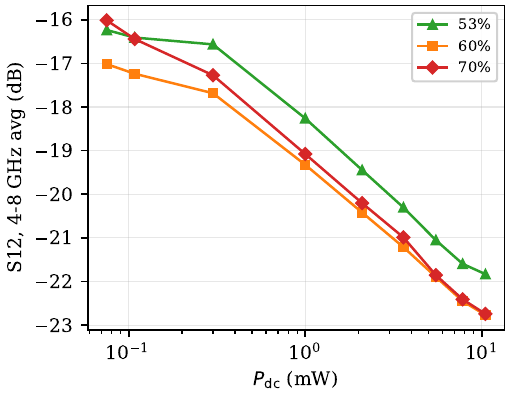}
  \caption{ HEMT averag $S_{12}$ over
           \SI{4}{}--\SI{8}{GHz} versus $P_\mathrm{dc}$
           for the three state-of-the-art devices.}
  \label{fig:s12}
\end{figure}

$P_1^\mathrm{th}$ is estimated directly from IQ blob statistics. For each bias point, the qubit is prepared in $\ket{0}$ and $P_1^\mathrm{th}$ is defined as the fraction of these $\ket{0}$-prepared shots classified as $\ket{1}$, $P_1^\mathrm{th} = P(1|0)$. As shown in Fig.~\ref{fig:thermal}, $P_1^\mathrm{th}$ decreases as
$P_\mathrm{dc}$ increases from its lowest tested value, consistent with improving reverse isolation of the HEMT at higher gain (quantified by the S-parameter $S_{12}$ shown in Fig.~\ref{fig:s12}); above $P_\mathrm{dc}\approx\SI{5.5}{mW}$ it rises again for several devices. The physical origin of this high-power upturn remains an open question. It is not explained by the $S_{12}$, which improves monotonically over
the entire bias range tested. \SI{4}{K} stage and the mixing-chamber
stage temperatures also do not track the increase in a consistent direction, ruling out simple self-heating of the shared cold stages. We therefore report the high-power fidelity decline as a real, device-dependent effect that is not explained by any of the candidate mechanisms we were able to test, rather than attribute it to a specific cause.

% ===============================================================
\section{Discussion}
\label{sec:discussion}
% ===============================================================

\subsection{Qubit-Based vs.\ Y-Factor Benchmarking}

LNA Y-factor noise temperature $\Tn$ from~\cite{ouryFactor} shows similar value for the 60\% and 70\% devices, whereas the qubit-based measurement ranks the 60\%
device gives the highest SNR and assignment fidelity.
This is consistent with the $T_{d}$ reported in~\cite{ouryFactor} (Figures 6 and 8), which is lowest for the 60\% device.

\begin{figure}[!t]
  \centering
  \includegraphics[width=\columnwidth]{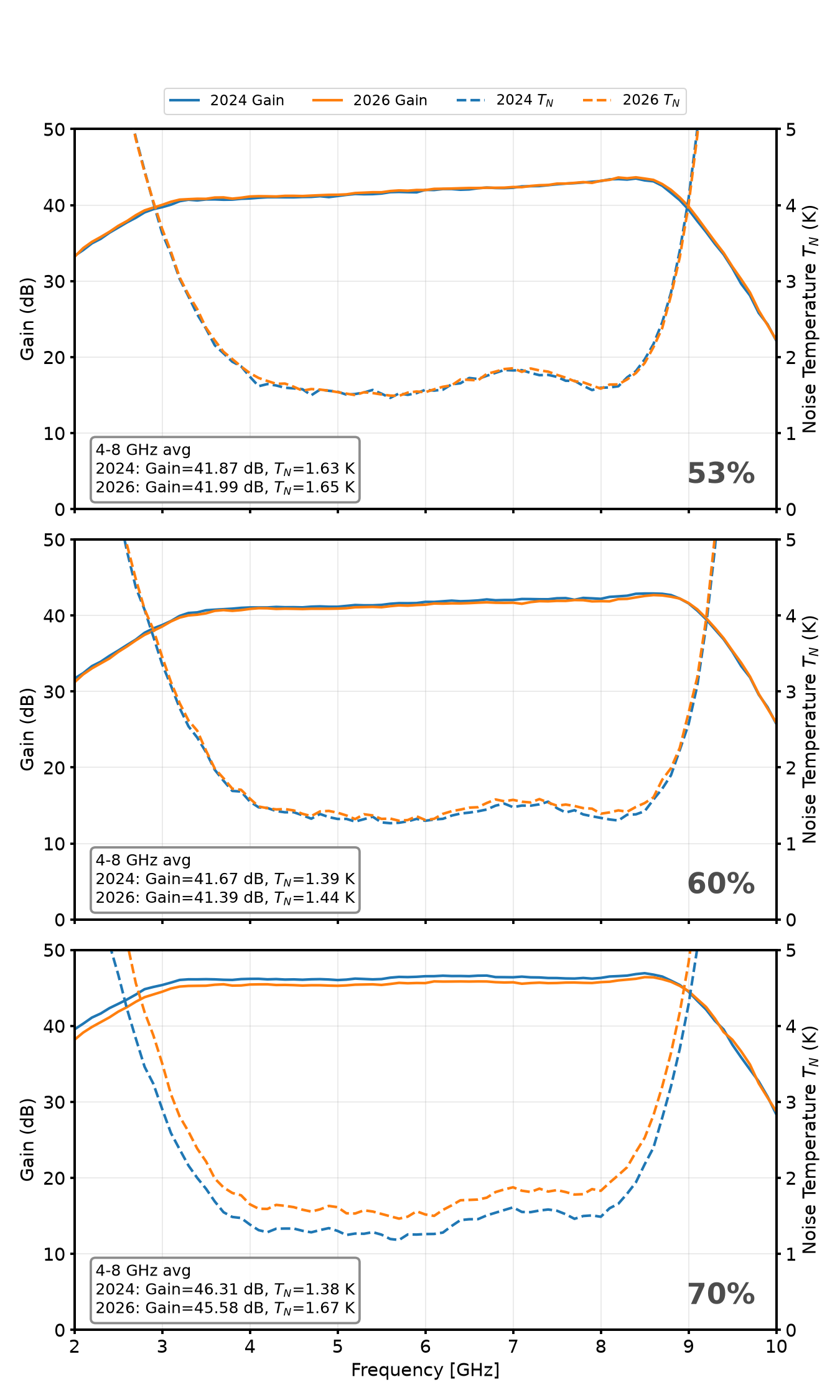}
  \caption{Gain and noise of the LNAs versus frequency at $V_{DS}=\SI{0.7}{V}$,
           $I_D=\SI{15}{mA}$, comparing the 2024
           characterization~\cite{ouryFactor} (blue) against a
           2026 measurement performed alongside this work
           (orange), for the 53\%, 60\%, and 70\% devices.}
  \label{fig:drift_check}
\end{figure}

Because the Y-factor characterization in~\cite{ouryFactor} predates the qubit measurements reported here, it is possible that one or more devices' noise performance has since changed. To check for this possibility, we repeated the gain and noise measurement for all three state-of-the-art devices at
$V_{ds}=\SI{0.7}{V}$, $I_D=\SI{15}{mA}$ and compared it against the characterization from the year 2024 ~\cite{ouryFactor}
(Fig.~\ref{fig:drift_check}). Averaged over 4–8 GHz, the 53\% device changed from Gain=41.87 dB, $T_{N}$=1.63 K (2024) to 41.99 dB, 1.65 K (2026), and the 60\% device from 41.67 dB, 1.39 K to 41.39 dB, 1.44 K: both changes are small and consistent with measurement-to-measurement scatter. The 70\% device, by contrast, changed from 46.31 dB, 1.38 K to 45.58 dB, 1.67 K, a decrease in gain of 0.73 dB together with an increase in noise temperature of 0.29 K, indicating that this device has degraded since its original characterization in both gain and $T_{N}$. Because the qubit measurements in this work were performed after this repeat characterization, the 70\% device's true noise
performance at the time of the qubit measurements is somewhat worse than the value reported in Table~\ref{tab:hemt_params}.

In addition, the qubit-based approach reveals a phenomenon
invisible to Y-factor. At high drain currents, we observe an increase in  $P_1^\mathrm{th}$ (Fig.~\ref{fig:thermal}) and a corresponding decline in $\fid_a$ for several devices. As discussed in Section~\ref{sec:bias_sweep}, this effect is not explained by the amplifier's reverse isolation, nor by 4\,K-stage
or mixing-chamber thermometry recorded during the same
measurements. It is not captured by Y-factor, which uses a matched resistive load rather than a qubit coupled to a superconducting resonator, and its physical origin remains an open question.

\subsection{Implications for Multi-Qubit Systems}

The results in Table~\ref{tab:pmin} have direct implications for
the design of large-scale, multiplexed-readout
processors~\cite{heinsoo2018multiplexed}.
At \SI{4}{K}, a typical pulse-tube refrigerator provides
\SI{1.5}{\watt} of cooling power~\cite{krinner2019cryogenic}.
Assuming an \SI{85}{\%} fidelity requirement and using the
optimal device (60\%), this budget accommodates
$\sim$1500 simultaneous readout chains at minimum bias
(Table~\ref{tab:pmin}). With frequency multiplexing per chain
(e.g.\ $\sim$7 qubits per chain, as demonstrated
in~\cite{heinsoo2018multiplexed}), this could support up to
$\sim$10,000 qubits, compared to only $\sim$1000 qubits at the
nominal bias of the least efficient device.

\subsection{Outlook: Combining with Fast Readout Protocols}

A demonstration using a similar setup with the commercial HEMT LNA achieved \SI{99.5}{\percent} assignment fidelity for two-state readout and \SI{96.9}{\percent} for three-state readout, also without a quantum-limited amplifier~\cite{chen2023readout}. It used a \SI{140}{ns} readout
time together with a shelving technique that suppresses the
contribution of decay error during readout to a residual
\SI{0.03}{\percent}, plus two-tone excitation and
machine-learning-assisted state discrimination.
That work notes that its fidelity could be further improved by a better amplifier, indicating that combining the better device demonstrated here with a shorter-integration-time, shelving-type protocol could in principle bring the best assignment fidelity closer to the \SI{99.5}{\percent} benchmark of~\cite{chen2023readout}.

% ===============================================================
\section{Conclusion}
\label{sec:conclusion}
% ===============================================================

We have presented a qubit-in-the-loop methodology for benchmarking
cryogenic HEMT LNAs in the context of superconducting qubit readout.
By analyzing the statistics of single-shot IQ histograms, we have demonstrated a sensitive and practically relevant figure of merit for HEMT LNA evaluation that is complementary to conventional Y-factor measurements.

Applying this methodology to three HEMT channel indium content LNAs over a wide range of dc power bias, we find that the 60\% device outperforms both the 53\% and 70\% devices, which is consistent with the $T_{d}$ extracted from the Y-factor LNA noise characterization. In addition, significant power reductions (more than 20 $\times$) are achievable with only modest fidelity penalties for sufficiently high-$\Tone$ qubits and moderate readout integration times. Moreover, assignment fidelity does not simply saturate with increasing power. For several devices it declines above a device-dependent power bias, tracking an increase in $P_1^\mathrm{th}$. This sets an independent upper bound on useful drain current in addition to the power-saving considerations above.

These results provide practical guidelines for HEMT LNA bias selection in
power-constrained multi-qubit systems, and establish qubit-based IQ
histogram benchmarking as a valuable addition to the cryogenic
amplifier characterization toolkit.

% ===============================================================
% Acknowledgments
% ===============================================================
\section*{Acknowledgment}

The author thanks Janka Biznárová from ConScience AB for providing the qubit chip; Tong Liu, Eleftherios Moschandreou, Stefan Hill, Martin Ahindura, Adilet Tuleuov, Pontus Vikstål, and Mårten Skogh from Chalmers Next Labs AB for the Tergite control software stack and for their discussions and support; and Jan Grahn from Chalmers MC2 TML and Johan Bergsten from Low Noise Factory AB for supporting and performing the LNA noise check. This work was supported by the Knut and Alice Wallenberg Foundation through the Wallenberg Centre for Quantum Technology (WACQT).

% ===============================================================
% References
% ===============================================================
\bibliographystyle{IEEEtran}
\bibliography{references}

\end{document}